\documentclass[aps,prb,twocolumn,floatfix,amsmath,amssymb]{revtex4}
\usepackage[final]{graphicx}
\usepackage{bm}
\usepackage{afterpage}
\usepackage{color}
\usepackage{comment}
\usepackage[colorlinks=true,urlcolor=blue,linkcolor=blue,citecolor=blue]{hyperref}
\usepackage{amssymb}

\emergencystretch=\maxdimen
\newcommand{\bk}{{\bf k}}

\begin{document}

\title{  Quantum Monte Carlo study of topological phases on a spin analogue of
Benalcazar-Bernevig-Hughes model }

\author{Jiaojiao Guo}
\affiliation{Department of Physics, Key Laboratory of Micro-Nano Measurement-Manipulation and Physics (Ministry of Education), Beihang University,
Beijing, 100191, China}

\author{Junsong Sun}
\affiliation{Department of Physics, Key Laboratory of Micro-Nano Measurement-Manipulation and Physics (Ministry of Education), Beihang University,
Beijing, 100191, China}

\author{Xingchuan Zhu}
\affiliation{Department of Physics, Beijing Normal University, Beijing, 100875, China}

\author{Chang-An Li}
\email{changan@connect.hku.hk}

\affiliation{Institute for Theoretical Physics and Astrophysics, University
of Würzburg, D-97074 Würzburg, Germnany}

\author{Huaiming Guo}
\email{hmguo@buaa.edu.cn}
\affiliation{Department of Physics, Key Laboratory of Micro-Nano Measurement-Manipulation and Physics (Ministry of Education), Beihang University,
Beijing, 100191,China}

\author{Shiping Feng}
\affiliation{ Department of Physics,  Beijing Normal University, Beijing, 100875, China}

\begin{abstract}
We study the higher-order topological spin phases based on a spin
analogue of Benalcazar-Bernevig-Hughes model in two dimensions using
large-scale quantum Monte Carlo simulations. A continuous N\'eel-valence
bond solid quantum phase transition is revealed by tuning the ratio between dimerized spin couplings, namely, the weak and strong
exchange couplings. Through the finite-size scaling analysis, we identify
the phase critical points, and consequently,
map out the full phase diagrams in related parameter spaces. Particularly,
we find that the valence bond solid phase can be a higher-order topological
spin phase, which has a gap for spin excitations in the bulk while
demonstrates characteristic gapless spin modes at corners of open
lattices. We further discuss the connection between the higher-order
topological spin phases and the electronic correlated higher-order
phases, and find both of them possess gapless spin corner modes that
are protected by higher-order topology. Our result exemplifies higher-order
physics in the correlated spin systems and will contribute to further
understandings of the many-body higher-order topological phenomena.
\end{abstract}


\maketitle

\section{Introduction}

Recently, a new family of higher-order topological insulators (HOTIs)
have attracted extensive attentions due to their novel and fundamental
physics \cite{benalcazar2017a,benalcazar2017,zhidasong2017,langbehn2017,khalaf2018,Schindler_2018,slager2015,bitan2019}.
An $n$th-order HOTI, like its conventional cousins, has gapless excitations
but at even lower $(d-n)$-dimensional boundaries that are protected
by higher-order topological invariants defined in the $d$-dimensional
bulk. For instance, the electric quadrupole insulator, proposed by Benalcazar et al., features zero-energy modes at the corners protected  by a quantized electric quadrupole moment in the bulk \cite{benalcazar2017a,benalcazar2017}. This new concept of higher-order topological protection on states
of matter has been immediately generalized to semimetals\cite{ezawa2018,maolin_2018},
superconductors\cite{zhongboyan2018,wanngqiyue2018}, and even non-Hermitian
systems\cite{kunst2018,taoliu2019,Ghatak2019}, constituting rich
higher-order topological phases. From the experimental aspect, HOTIs
have been reported in several platforms, such as the electric circuits\cite{imhof_2018,serra_2019,ezawa_2018},
microwave resonators\cite{peterson_2018}, classical mechanical metamaterials\cite{ma_2019,haiyanfan_2019},
photonic and phononic crystals\cite{zilberberg_2018,noh_2018,xue_2019,ni_2019,xie_2018,zhang2019second,biyexie_2019,el2019corner,xiaodongchen_2019,mittal2019photonic},
although they are difficult to be found in electric systems \cite{schindler_2018a,bingliu_2019,xianleisheng2019,xianleisheng2020}.

Up to date, however, the main attention paid on the higher-order topological
phases is within the framework of topological band theory, their closely
related counterparts taking account the strong correlation effects
are still less known. Introducing correlations will unexpectedly bring
new and different properties to the system. For instance, it is found
that the quantized electric quadrupole insulators are robust against
weak interactions but are driven to antiferromagnetic insulators by
strong enough correlations \cite{zylu2020}. Besides, the electron
correlations will generalize the bulk-boundary correspondence, i.e.,
there emerge gapless corner modes only in spin excitations while the
single-particle excitations remain gapped \cite{hatsugai2019}. The
bosonic counterparts of HOTIs have also been proposed in a two-dimensional
superlattice Bose-Hubbard model and dimerized spin systems\cite{pollmann2020,hughes2019,yizhiyou2018},
and the representative signatures such as fractional corner charges
are explicitly demonstrated using the density matrix renormalization group\cite{pollmann2020}.

To characterize the topological properties in system with strong correlations,
there are several proposals for the possible many-body topological invariant.
One recipe is the Green's function formalism that is widely used in
quantum Monte Carlo (QMC) simulations. Based on the zero-frequency
Green's function, the nested Wilson loop method can be directly applied
to obtain the many-body topological invariants \cite{zylu2020,zhongwang2012,Wang_2013}.
Another method directly extends the charge polarization of one-dimensional
(1D) systems\cite{resta_1998} to many-body order parameters for bulk
multipole moments in 2D and 3D systems, which has been verified to
give the correct phase diagrams of topological multipole insulators\cite{wheeler_2019,byungminkang_2019}
despite its finite defects \cite{watanabe_2019}. Owing to its real
space nature, it also allows the investigation of disorder effects
in these electric multipole insulators\cite{changanli2020,changanli2020a,bitan2020}.
Furthermore, the $\mathbb{Z}_{Q}$ Berry phase may be an alternative
topological invariant to characterize the HOTIs and the correlated
ones\cite{araki_2020}.
We note the latter two may encounter finite-size difficulties
when implementing many-body calculations using the exact diagonalization
method that is available only on very limited lattice sizes\cite{wheeler_2019,araki_2020}.

As we know, numerical techniques, such as QMC method, contribute significantly
to the understanding of exotic properties in the strongly correlated
systems. The main reason is the absence of exact solutions and controlled
analytic approximations for the strongly interacting systems. Although
remarkable advances have been made in studying the correlated higher-order
topological phases so far, a systematical understanding is still incomplete.
Indeed, we note that the higher-order topological spin phases are
even largely unexplored by unbiased large-scale numerical simulations.

In this work, we employ a sign-problem-free QMC method to study the
higher-order topological properties in a spin-model analogue of the
Benalcazar-Bernevig-Hughes (BBH) model. We first treat the spin BBH
model with Heisenberg couplings using linear spin wave theory (LSWT).
While the N\'eel-valence band solid (VBS)\cite{vbs} transition can be qualitatively
described by LSWT, the critical values are greatly underestimated.
Then we determine the precise transition points using finite-size
scaling of dimensionless quantities characterizing the antiferromagnetism
(AFM) obtained by the QMC simulations. The phase diagrams in the $(g,\Delta)$
and $(J_{1x},J_{1y})$ planes are mapped out, in which the VBS phases
with $J_{1x(y)}<J_{2}$ are identified to be spin higher-order topological
phase (SHOTP), which are analogues of the HOTIs in
the BBH model, and have gapless corner modes in the spin excitations. The gapless
spin corner modes are demonstrated by applying a perpendicular magnetic
fields to open lattice. In the presence of a spin gap, only four free
corner spins can be aligned by the magnetic field, generating a quantized
magnetization $M_{z}=2$. Besides, the magnetization mainly distributes
near the corners, confirming the gapless corner spins cause the magnetization
under the magnetic field. We finally discuss the connection between
the SHOTPs and the electronic correlated HOTIs.

This paper is organized as follows. Section II introduces the precise
model we will investigate, along with our computational methodology.
Section III presents LSWT of the spin BBH model with Heisenberg
couplings. Section IV uses QMC simulations to study the bulk properties
of the spin models. Section V demonstrates the spin corner states. Section VI shows the results of  higher-order topological invariants.
Section VII discusses the connection to the fermionic BBH-Hubbard model,
and is followed by some further discussion and conclusion in Section
VIII.

\section{The model and method}

\begin{figure}[htbp]
\centering \includegraphics[width=4.5cm]{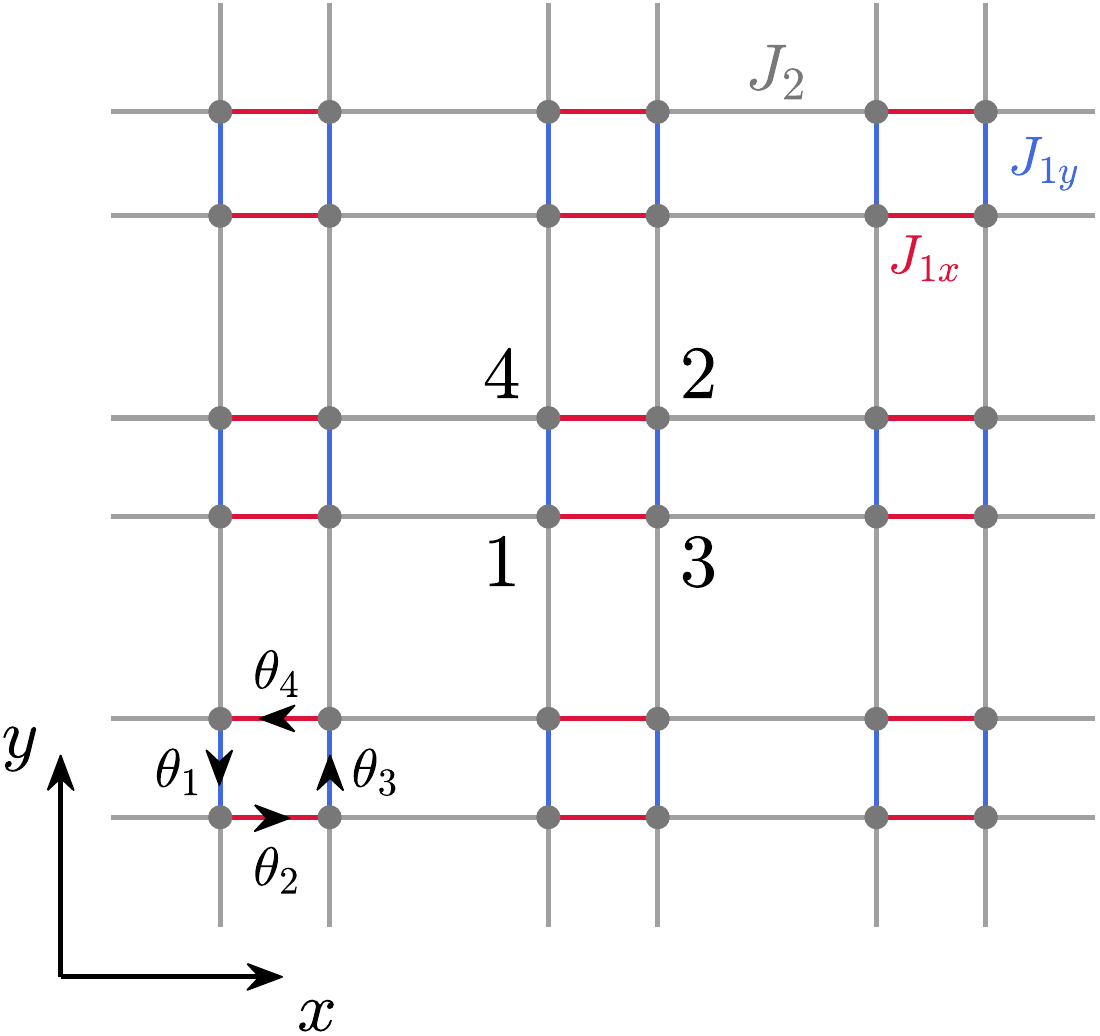} \caption{ Schematics of the spin analogue of the BBH model described by the Hamiltonian in Eq.(1). Each unit cell contains four sites. Short lines represent intra-cell weak exchange couplings $J_{1x}, J_{1y}$, and all units are connected by long lines corresponding to strong couplings $J_2$. The left-bottom plaquette exhibits the twist parameters for the $\mathbb{Z}_4$ Berry phase.}
\label{fig1}
\end{figure}

We consider the following dimerized spin Hamiltonian on a square lattice with four spin-$1/2$ degrees of freedom per unit cell,
\begin{eqnarray}\label{eq1}
\hat{H}_0=\sum_{{\bf r}}&[& J_{1,x}(\hat{h}_{{\bf r},1;{\bf r},3}+\hat{h}_{{\bf r},2;{\bf r},4}) \\ \nonumber
&+&J_{1,y}(\hat{h}_{{\bf r},1;{\bf r},4}+\hat{h}_{{\bf r},2;{\bf r},3})  \\ \nonumber
&+&J_{2,x}(\hat{h}_{{\bf r},3;{\bf r}+\hat{x},1}+\hat{h}_{{\bf r},2;{\bf r}+\hat{x},4}) \\ \nonumber
&+&J_{2,y}(\hat{h}_{{\bf r},4;{\bf r}+\hat{y},1}+\hat{h}_{{\bf r},2;{\bf r}+\hat{y},3}) ]
\end{eqnarray}
where $\hat{h}_{{\bf r},i;{\bf r'},j}=[S^{x}_{{\bf r},i}S^{x}_{{\bf r'},j}+S^{y}_{{\bf r},i}S^{y}_{{\bf r'},j}+\Delta S^{z}_{{\bf r},i}S^{z}_{{\bf r'},j}]$ with $i,j=1,2,3,4$ marking the site in the unit cell, and $\Delta$ a dimensionless parameter characterizing
the anisotropy of the exchange couplings. $S^{\gamma}_{{\bf r},i}$ is spin-$1/2$ operator on the site $i$ of the unit cell at ${\bf r}$, which obeys commutation relations, $\left[S^{\alpha}_{{\bf r},i},S^{\beta}_{{\bf r},j}\right]=i\hbar \varepsilon _{\alpha\beta\gamma}S^{
\gamma}_{{\bf r},i}\delta_{ij}\delta_{{\bf r},{\bf r'}}$ with $\varepsilon _{\alpha\beta\gamma}$ the Levi-Civita symbol and $\alpha,\beta,\gamma= x,y,z$ representing spin direction. $J_{1,x},J_{1,y} (J_{2,x},J_{2,y})$ are intra(inter)-cell exchange antiferromagnetic (AF) couplings. The model with $\Delta=0(1)$ corresponds to a dimerized XY(Heisenberg) spin system. By tuning the anisotropy $\Delta$, the topological property of the XXZ spin model can also be explored. Throughout the manuscript, we set $J_{2,x}=J_{2,y}=J_2=1$ as the energy scale.

In the following discussions, we employ the approach of stochastic series expansion (SSE) QMC method~\cite{sandvik2002,syljuasen2003} with directed loop updates to study the model in Eq.(1). The SSE method expands the partition function in power series and the trace is written as a sum of diagonal matrix elements. The directed loop updates make the simulation very efficient~\cite{Bauer2011,fabien2005,pollet2004}. Our simulations are on square lattices with the total number of sites $N=4 L \times L$ with $L$ the linear size. There are no approximations
causing systematic errors, and the discrete configuration space can be sampled without floating
point operations. The temperature is set to be low enough to obtain the ground-state properties. For such spin systems on bipartite lattice, the notorious sign problem in the QMC approach can be avoided.

\section{The linear spin wave theory}

\begin{figure}[htbp]
\centering \includegraphics[width=7.5cm]{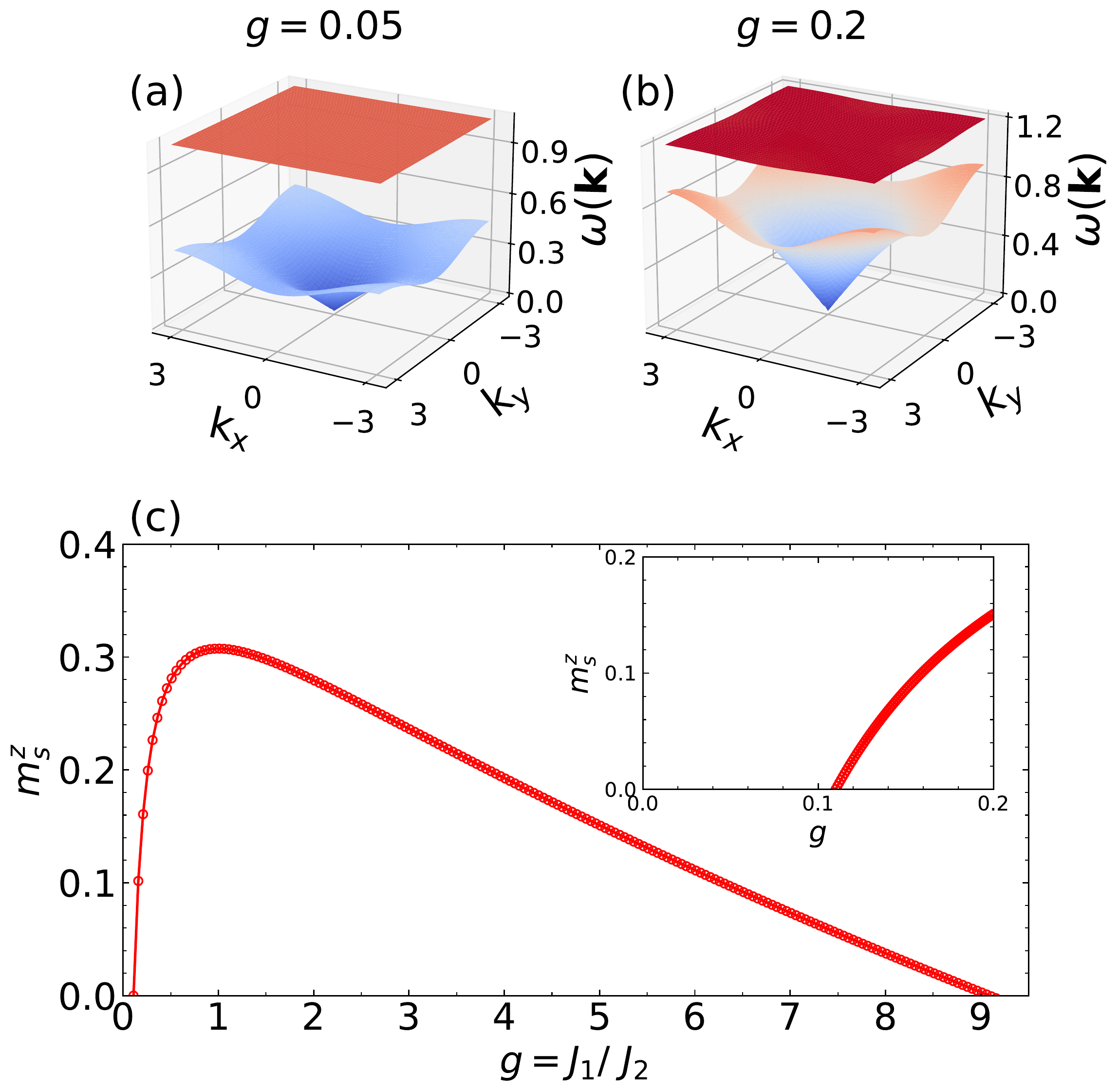} \caption{The magnon spectrums at (a) $J_1/J_2=0.05$ and (b) $J_1/J_2=0.2$, corresponding to the cases without and with long-range AF orders, respectively. (c) The AF order
parameter from the LSWT analysis, Eq. (8). Within
LSWT, long-range AF order appears above $g_c =0.11$. Here the results are based on the Heisenberg model with $\Delta=1$.}
\label{fig2}
\end{figure}
Let us first get some insights on the properties of the model Eq. (1) by LSWT analysis.
The Heisenberg model with $\Delta=1$ described by Eq.(1) can be treated within LSWT by replacing the spin operators by bosonic ones via Holstein-Primakoff (HP) transformation\cite{hptransformation}. The transformation on sublattice $1,2$ (the spin is in the positive $z$-direction) is defined as
\begin{align}\label{eq2}
S^+_{1(2),i}&=a_{i,1(2)}, S^-_{1(2),i}=a^{\dagger}_{i,1(2)},\\ \nonumber
S^z_{1(2),i}&=\frac{1}{2}-a^{\dagger}_{i,1(2)}a_{i,1(2)}.
\end{align}
On sublattice $3,4$ (the spin is in the negative $z$-direction), the spin operators are defined as
\begin{align}\label{eq3}
S^+_{3(4),i}&=b^{\dagger}_{i,3(4)},S^-_{3(4),i}= b_{i,3(4)},\\ \nonumber
S^z_{3(4),i}&=b^{\dagger}_{i,3(4)} b_{i,3(4)}-\frac{1}{2}.
\end{align}
Then the bosonic tight binding Hamiltonian becomes $H=\sum_{\bf k}X^{\dagger}_{\bf k}{\cal{H}}({\bf k})X_{\bf k}$, where $X_{\bf k}=\left(a_{1\mathit{\mathbf{k}}}^{\dagger } \;\;a_{2\mathit{\mathbf{k}}}^{\dagger } \;\;b_{3\mathit{\mathbf{k}}} \;\;b_{4\mathit{\mathbf{k}}} \;\right)^{T}$ is the basis, and
\begin{equation}\label{eq4}
\cal{H}({\bf k})= \left\lbrack \begin{array}{cccc}
C_0  & 0 & \gamma_{1\mathit{\mathbf{k}}}  & \gamma_{2\mathit{\mathbf{k}}} \\
0 & C_0  & \gamma_{2\mathit{\mathbf{k}}}^{*}  & \gamma_{1\mathit{\mathbf{k}}}^{*} \\
\gamma^{*}_{1\mathit{\mathbf{k}}}  & \gamma_{2\mathit{\mathbf{k}}}  & C_0  & 0\\
\gamma_{2\mathit{\mathbf{k}}}^{* }  & \gamma_{1\mathit{\mathbf{k}}}  & 0 & C_0
\end{array}\right\rbrack,
\end{equation}
with
\begin{align}
\nonumber
C_0 &=\frac{1}{2}\left(J_{1x} +J_{1y}\right) +J_{2} \\ \nonumber
\gamma_{1\mathit{\mathbf{k}}}&=\frac{1}{2}\left(J_{1y} +J_{2} e^{-\textrm{i}k_y} \right)\\  \nonumber
\gamma_{2\mathit{\mathbf{k}}}&=\frac{1}{2}\left(J_{1x} +J_{2} e^{-\textrm{i}k_x} \right)\\ \nonumber
\end{align}

The Hamiltonian ${\cal H}({\bf k})$ in Eq.(4) can be diagnalized
using Bogliubov transformation\cite{2015Linear}. The spin wave spectrum
contains two branches, each of which is two-fold degenerate (see Appendix
A for the analytical form). The lower energy spin-wave excitations
in our model display a linear spectum for small $k$, which is consistent
with their AFM nature. For the $C_{4}$ symmetric case, we let $J_{1x}=J_{1y}=J_1$, and define the ratio $g=J_1/J_2$. The magnon spectrums
at $g=0.05,0.2$ are shown in Fig.2(a) and (b). Although
the spectrums are gapless, there is a gap between the two branches,
and its size monotonously decreases as $J_{1}$ is strengthened. The
gap vanishes at $J_{1}=J_{2}$ when the system becomes 2D AF Heisenberg
model. The resulting gapless spectrum with two branches is connected
to the normal one under a two-site unit cell by a folding of the Brillouin zone.

The AFM staggered order parameter defined as
\begin{equation}\label{eq5}
  m_s^{z} =\frac{1}{N}\left(\sum_{i\in (1,2)} \left\langle S_i^z \right\rangle -\sum_{i\in (3,4)} \left\langle S_i^z \right\rangle \right)
\end{equation}
is obtained in LSWT, where $i\in (1,2)$ denotes those sites which are spin up sublattice sites, and $i\in (3,4)$ denotes those sites which are spin down sublattice sites. Writing $\left\langle S_i^z \right\rangle$ in terms of HP operators, we have
\begin{eqnarray}\label{eq6}
&m^{z}_s& = \frac{1}{N}\left(\sum_{i\in (1,2)} \left\langle S-a_i^{\dagger } a_i \right\rangle -\sum_{i\in (3,4)} \left\langle b_i^{\dagger } b_i -S\right\rangle \right) \\ \nonumber
&=& S+\frac{1}{2}-\frac{1}{N}\left( \sum_{\bk}\sum_{i=1}^{4} \left\langle a_{i{\bk}}^{\dagger } a_{i{\bk}} \right\rangle \right),
\end{eqnarray}
where $S$ is the spin.

We then use the Bogliubov transformation to write the basis $X$ in terms of the diagonal basis $Y=\left(\alpha_{1\mathit{\mathbf{k}}}\;\;\alpha_{2\mathit{\mathbf{k}}} \;\;\beta_{3\mathit{\mathbf{k}}}^{\dagger}\;\;\beta_{4\mathit{\mathbf{k}}}^{\dagger} \;\right)^{T}$: $X=BY$. The Bogliubov transformation matrix $B$ satisfies $Bs_zB^{\dagger}=s_z$, with
\begin{equation}\label{eq7}
  s_z=\left\lbrack \begin{array}{cccc}
1 & 0 & 0 & 0\\
0 & 1 & 0 & 0\\
0 & 0 & -1 & 0\\
0 & 0 & 0 & -1
\end{array}\right\rbrack.
\end{equation}
At zero temperature, only the terms containing $\left\langle \beta_{3\mathit{\mathbf{k}}}\beta_{3\mathit{\mathbf{k}}}^{\dagger } \right\rangle =\left\langle \beta_{4\mathit{\mathbf{k}}}\beta_{4\mathit{\mathbf{k}}}^{\dagger } \right\rangle$ are nonzero $(=1)$. Hence the staggered order parameter is
\begin{equation}\label{eq8}
  m_s^{z}=1 - \frac{1}{N}\sum_{\mathbf{k}}((B^{\dagger}B)_{33} + (B^{\dagger}B)_{44})
\end{equation}
where $(B^{\dagger}B)_{ii}$ denotes the $i$-th element of the matrix $B^{\dagger}B$.
Figure 2(c) shows $m_s^z$ in Eq.(8) as a function of $g$. LSWT greatly underestimates the transition point, and it predicts the N\'eel-VBS transition at $g_c=0.11$. The periodic lattice in Fig.1 is invariant under interchanging $J_1$ and $J_2$ due to the translation symmetry, indicating a duality between $g$ and $1/g$. Hence there is also a critical value in the large $g$ region, which is exactly $1/g_c$ with $g_c$ the small critical value. Although the gap size is not affected by the duality, the topological property of the VBS phase changes to be trivial when $J_1$ becomes stronger.

\section{QMC simulations of the bulk properties}
In this section, we investigate the bulk propertires of the higher-order
topological spin models using QMC simulations. The properties of the
system are controlled by the ratio $J_{1x}/J_{2}$ and $J_{1y}/J_{2}$,
and there exist several scenarios. To obtain some insights, let us
first discuss two special limits $J_{1x}=J_{1y}=1$ and $J_{1x}=J_{1y}=0$,
at which the solutions are well known or exactly solvable. In the
former limit, the Hamiltonian Eq. (1) is the XXZ spin model. The case
with $\Delta=1$ corresponds to an isotropic AF Heisenberg model,
which has a ground state with long-range AF order. In the latter limit,
the lattice is decoupled into isolated $2\times2$ plaquettes. The
ground-state energy is $E_{0}=-J_{2}(\Delta+\sqrt{\Delta^{2}+8})/2$,
which is separated from the first-excited state by a gap with the
size $-J_2-E_{0}$ (see Appendix B). While the ground state is in the total $S_{z}=0$
sector, the first-excited state has $S_{z}=\pm1$. Hence the low-energy
excitation is a spin one, which is gapped in the four-site spin chain.

For general cases $J_{1x}\neq0$ as well as $J_{1y}\neq0$, a N\'eel-VBS
quantum phase transition can happen by tuning the coupling ratio $J_{1x}/J_{2}$
or $J_{1y}/J_{2}$. The phase transition points can be precisely determined
using the SSE QMC by measuring the staggered magnetization $m_{s}^{z}$,
the spin stiffness $\rho_{s}$, and the uniform susceptibility $\chi$.
Here, the staggered magnetization $m_{s}^{z}$ is defined as\cite{wenzel_2008,sandvik2010computational,nvsenma_2018,ranxiaoyue_2019,jiangyifan_2012}
\begin{eqnarray}
m_{s}^{z}=\frac{1}{N}\sum_{i}^{N}S_{i}^{z}(-1)^{x_{i}+y_{i}},\label{eq1}
\end{eqnarray}
and its Binder ratio is $Q_{2}=\frac{\langle(m_{s}^{z})^{4}\rangle}{\langle(m_{s}^{z})^{2}\rangle^{2}}$\cite{binder_1981,binder_1984}.
The uniform susceptibility writes as\cite{wenzel_2008,sandvik2010computational,nvsenma_2018,ranxiaoyue_2019}
\begin{eqnarray}
\chi=\frac{\beta}{N}\langle(\sum_{i}^{N}S_{i}^{z})^{2}\rangle.\label{eq1}
\end{eqnarray}
In terms of SSE configurations, the spin stiffness can be obtained
by the expression\cite{wenzel_2008,sandvik2010computational,nvsenma_2018,ranxiaoyue_2019}
\begin{eqnarray}
\rho_{s}=\frac{3}{2\beta}(W_{x}^{2}+W_{y}^{2}),\label{eq1}
\end{eqnarray}
where $W_{\alpha}=(N_{\alpha}^{+}-N_{\alpha}^{-})/L_{\alpha}$ is
the winding number in the $\alpha$ ($x$ or $y$) direction, which takes integer values, and represents the times of spin transporting across the system; $N_{\alpha}^{+}(N_{\alpha}^{-})$ denotes the total number of operators
transporting spin in the positive (negative) direction\cite{nadd}. Due to the Lorentz symmetry, the correlation length is the same in the space and time directions, and hence the dynamic exponent is $z=1$\cite{lorenzsymmetry}. The temperature appears in the argument $L^z/\beta$ of the scaling
function, thus we
took $\beta=L$ to exclude the temperature dependence in the finite-size scaling. At the quantum critical point, the dimensionless quantities
$Q_{2},L\chi,L\rho_{s}$ are size-independent, and should cross for
different lattice sizes, from which we can extract the phase transition
points without knowing
the critical exponents.

\begin{figure}[htbp]
\centering \includegraphics[width=6.5cm]{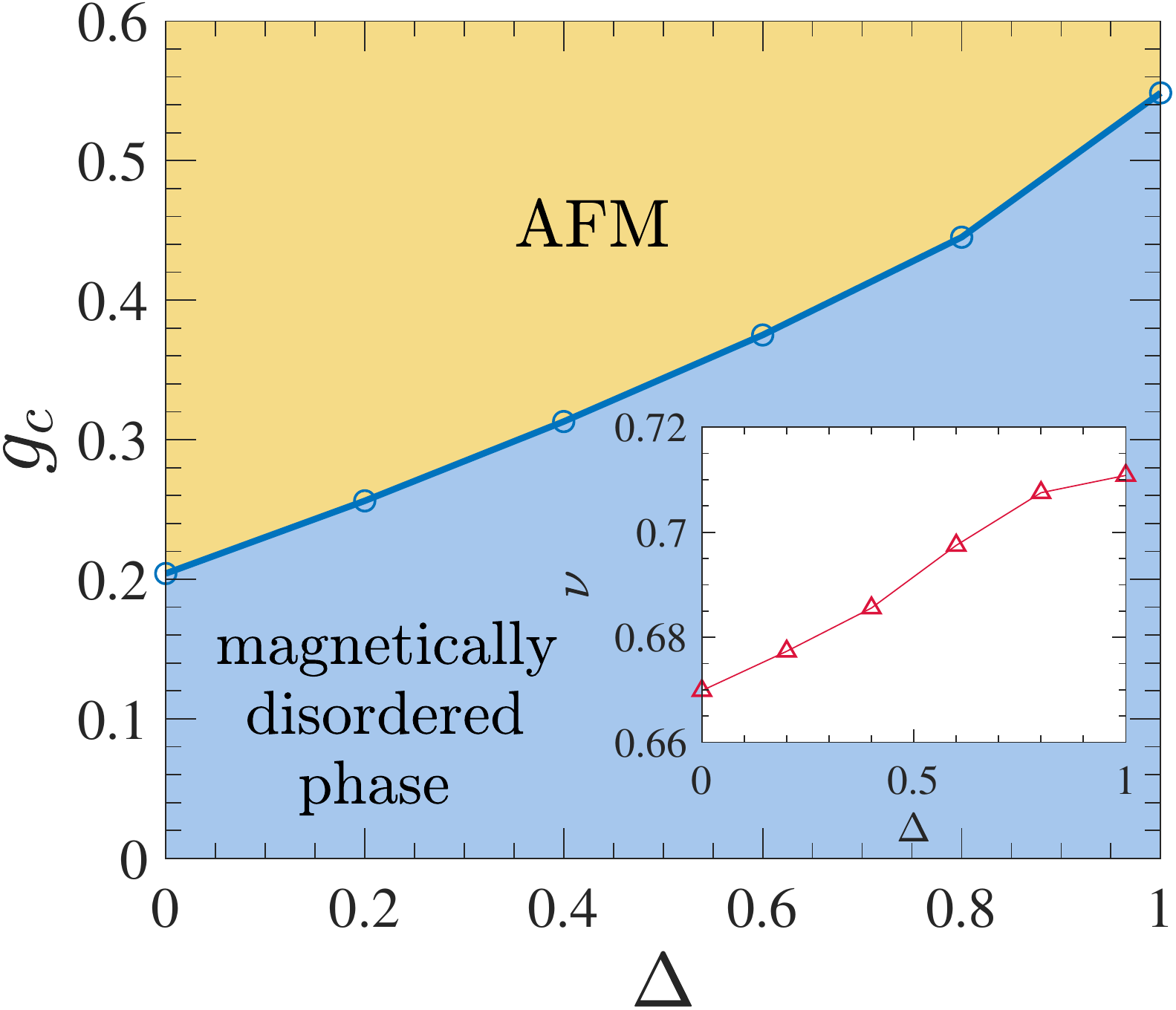} \caption{Phase diagram in the $(\Delta, g_c)$ plane for the $C_4$ symmetric ($J_{1x}=J_{1y}$) Hamiltonian in Eq.(1). Inset shows the correlation length critical exponent $\nu$ estimated from the best data collapse of the dimensionless quantities $Q_2, L\chi, L\rho_s$. Here the magnetically disordered phase is a VBS\cite{vbs}.}
\label{fig3}
\end{figure}
\begin{figure}[htbp]
\centering \includegraphics[width=7.5cm]{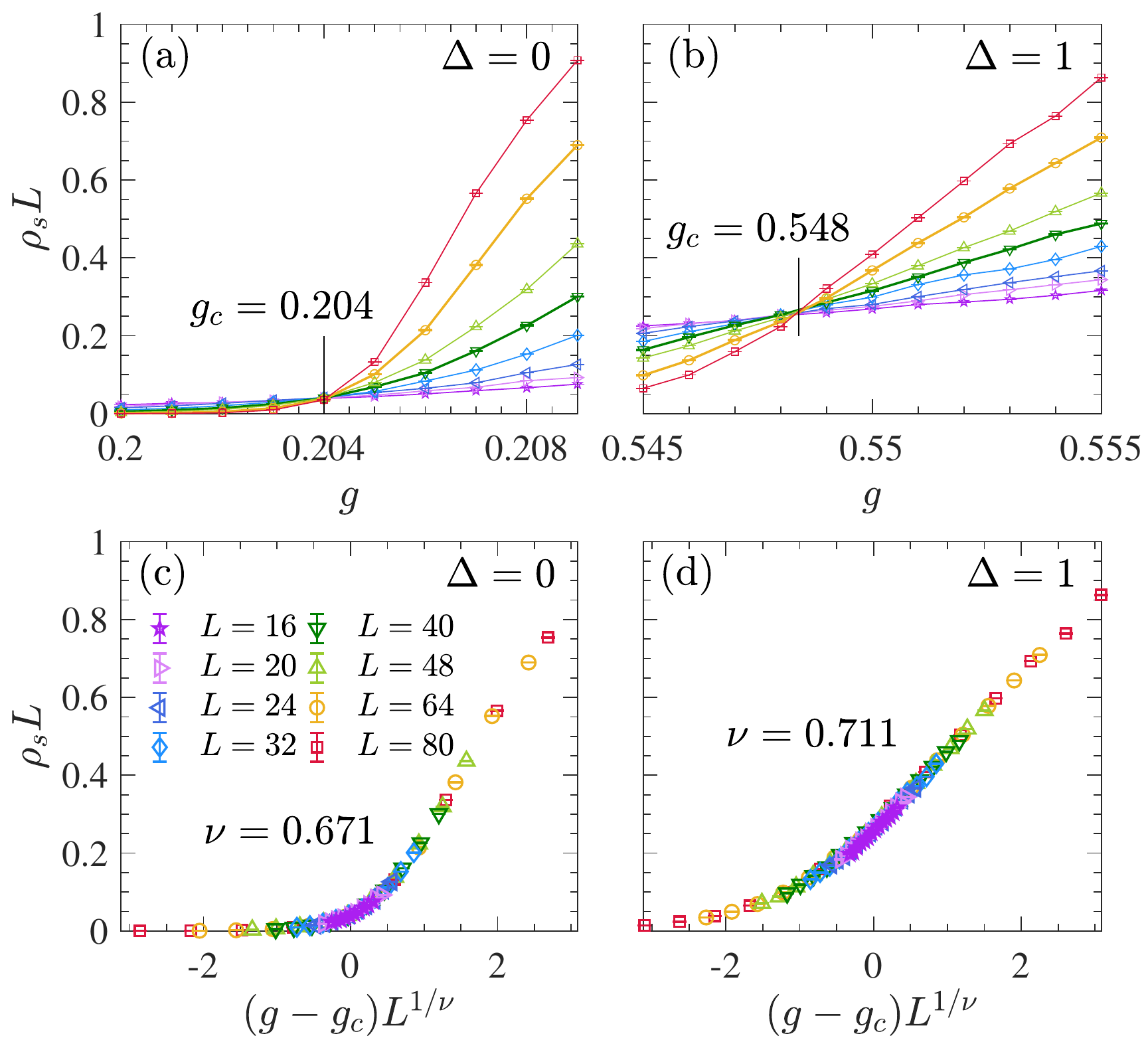} \caption{ $L\rho_s$ as a function of $g=J_1/J_2$ for the $C_4$ symmetric Hamiltonian: (a) $\Delta=0$; (b) $\Delta=1$. The universal crossings determine the transition points $g_c=0.204$ in (a) and $g_c=0.548$ in (b). (c) and (d) are the corresponding best scaling collapses of $L\rho_s$, giving the correlation length critical exponent $\nu=0.671$ for $\Delta=0$ and $\nu=0.711$ for $\Delta=1$, respectively. In all figures, the same lattice sizes are used, and the symbol-size correspondences are shown in (c).}
\label{fig4}
\end{figure}

Let us first consider the special case with $C_{4}$ symmetry, i.e., $J_{1x}=J_{1y}$ (the value is denoted as $J_{1}$). Figure \ref{fig3}
shows the critical point $g_{c}$ as a function of the anisotropy $\Delta$.
For $g<g_{c}$, the system is in the VBS phase, characterized by $m_{s}^{z}=\rho_{s}=\chi=0$.
The AF long-range order develops above $g_{c}$, where the values of the
staggered magnetization, the spin stiffness and the uniform susceptibility
become finite. Since the N\'eel-VBS phase transition is continuous, $g_{c}$ is determined by the crossings of the above dimensionless
quantities. As examples, we show $L\rho_{s}$ at $\Delta=0,1$ for
different lattice sizes in Fig.\ref{fig4}. Using the leading scaling
ansatz for a second-order phase transition,
\begin{eqnarray}\label{eq1}
L\rho_s=f[(g-g_c)L^{1/\nu}],
\end{eqnarray}
the correlation length critical exponent $\nu$ is determined by the best data collapse. As shown in the inset of Fig.\ref{fig3}, $\nu$  increases continuously with $\Delta$. The Heisenberg model with $\Delta=1$ belongs to the three-dimensional (3D) O(3) universality, and the obtained value of $\nu$ is consistent with the standard O(3) one of $\nu=0.7112(5)$\cite{vicari2002a,vicari2002}
For the XY model with $\Delta=0$, the universality class is the 3D XY one. The critical exponent we obtain is $\nu=0.671$, consistent with the accurate results in the literature\cite{vicari2002}.

\begin{figure}[htbp]
\centering \includegraphics[width=8.5cm]{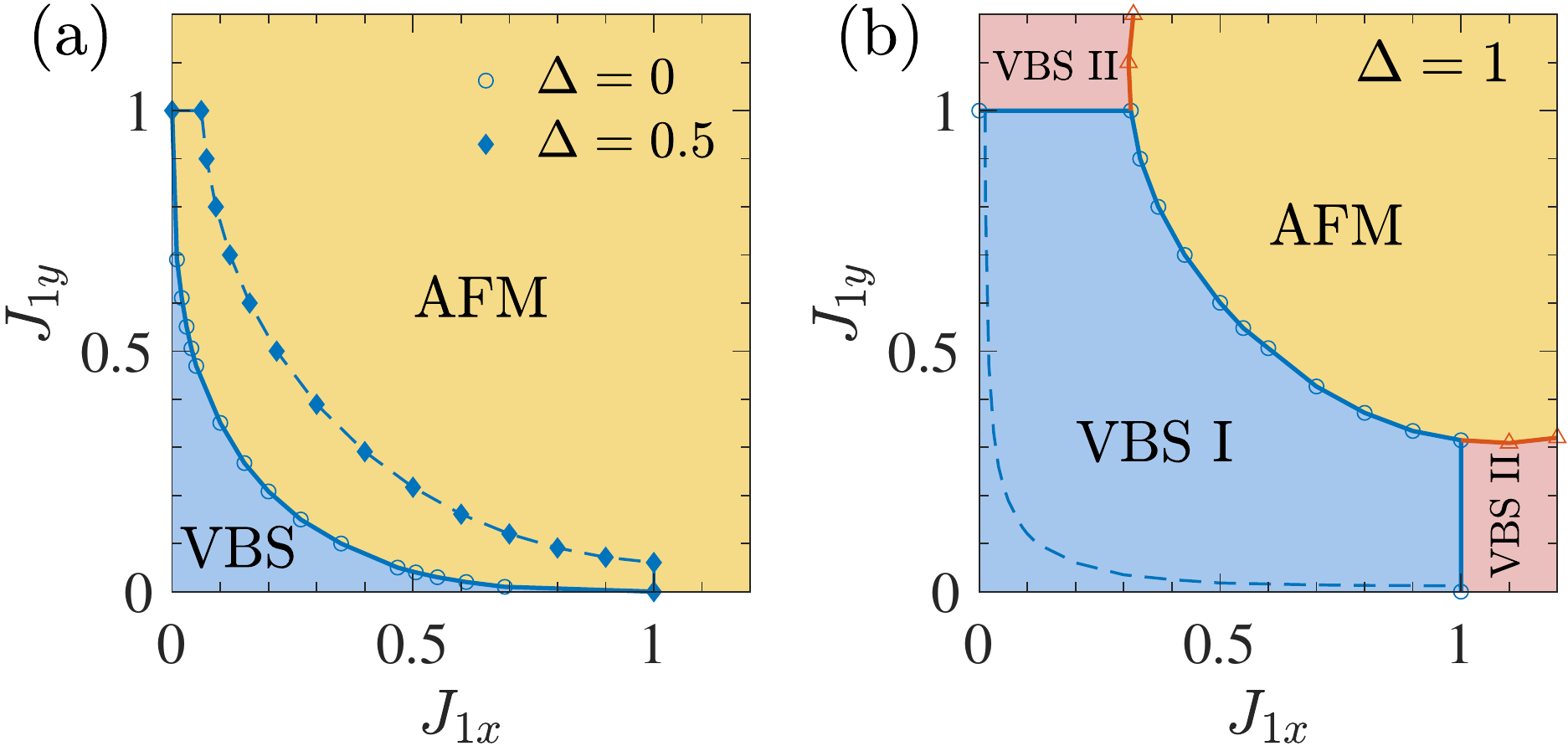} \caption{Phase diagrams in the $(J_{1x}, J_{1y})$ plane for the model in Eq.(1) with (a) $\Delta=0,0.5$; (b) $\Delta=1$. Dashed lines in (b) indicate the phase boundaries from LSWT. The VBS in (a) and VBS I in (b) are SHOTPs, which have bulk gapped spin excitations, and gapless spin corner states on open lattice. Here VBS I and VBS II are defined relatively to the boundary
of the geometry in Fig.1.}
\label{fig5}
\end{figure}

Next we map out the phase diagram for general cases in the $(J_{1x},J_{1y})$
plane. Here we only focus on the Heisenberg and XY exchange couplings,
and the phase diagrams for $0<\Delta<1$ are similar. The critical
points are determined more precisely using the best data collapses
with the accurate $\nu$ for the O(3) and XY universality classes.
The phase diagrams in Fig. \ref{fig5} consist of the AFM and VBS
phases. The region of VBS shrinks as $\Delta$ is decreased. In both
figures, the VBS phases with $J_{1x(y)}/J_{2}<1$ are SHOTPs,
which will be demonstrated in the following section. Here the SHOTPs survive only in part of the unit square. In contrast, the electronic
HOTIs of the BBH model are in the entire square region $|t_{1x(y)}/t_{2}|<1$
{[} $t_{1x(y)}(t_{2})$ denotes the hopping amplitude of the weak
(strong) bonds{]}.

Then we consider another limit $J_{1x}=0$ or $J_{1y}=0$, at which
the system becomes one-dimensional dimerized spin chain on the boundaries. Under the
Jordan-Wigner transformation\cite{coleman2015}
\begin{align}
S_{i,1}^{+} &=\alpha_{i}^{\dagger}\cdot e^{i \pi \sum_{j<i}\left(n_{i,\alpha}+n_{i,\beta}\right)} \\ \nonumber
S_{i,2}^{+} &=\beta_{i}^{\dagger}\cdot e^{i \pi \sum_{j<i}\left(n_{j,\alpha}+n_{j,\beta}\right)} e^{i \pi n_{i,\alpha}}, \\ \nonumber
S^{z}_{i,1(2)} &= n_{i,\alpha(\beta)}-1/2 \nonumber
\end{align}
the following 1D fermionic Hamiltonian is obtained,
\begin{align}\label{eq1}
H_{f}=\sum_{i}[&\frac{1}{2}(J_{1x(y)}\alpha_i^{\dagger}\beta_i+J_{2}\beta^{\dagger}_i\alpha_{i+1}+\textrm{H.c.})+ \\ \nonumber
&J_{1x(y)}\Delta(n_{i,\alpha}-\frac{1}{2})(n_{i,\beta}-\frac{1}{2})+ \\ \nonumber
&J_{2}\Delta(n_{i,\beta}-\frac{1}{2})(n_{i+1,\alpha}-\frac{1}{2})],
\end{align}
where $\alpha_i,\beta_i$ are annihilation operators of spinless fermions on the two sites of unit cell $i$, and $n_{i,\alpha}=\alpha^{\dagger}_i\alpha_i,n_{i,\beta}=\beta^{\dagger}_i\beta_i$ are the corresponding number operators. In the hopping term connecting the boundary, there is an additional phase $e^{i\pi n_{t}}$ ($n_t$ the total number of fermions), which may be $1$ or $-1$ depending on even or odd $n_t$. Since such a sign has no effect on the 1D boundary mode, it does not affect the topological property of the system, and we omit it in the above Hamiltonian.
For the XY model, the mapped fermionic Hamiltonian is the noninteracting Su-Schrieffer-Heeger (SSH) model\cite{wupeisu1979}, which has a topological phase transition at $J_{1x(y)}=J_2$. The 1D dimerized Heisenberg model corresponds to an interacting SSH model with dimerized nearest-neighbor repulsions. It has been shown that an AFM transition occurs at $J_{1x(y)}=J_2$\cite{mrigol2011,mrigol2011a}. We calculate the 1D topological invariant, i.e., the Zak phase, which is $\gamma=1$ for $J_{1x(y)}<J_2$ and $\gamma=0$ for $J_{1x(y)}>J_2$ (see Appendix C). This unambiguously shows the topological phase transition happens exactly at the uniform point, which determines the on-axis points of the phase boundary in Fig.\ref{fig5}.

It is noted that the transition line between VBS I and VBS II is a straight line. In this region, the dimensionless quantities, such as $L\rho_s$ $L\chi$, do not exhibit universal crossings as $J_{1x} (J_{1y})$ varies at fixed $J_{1y} (J_{1x})$, and instead their values continuously decrease with increasing $L$. These imply the absence of a N\'eel-VBS transition, and the system keeps in the VBS phase. Since there is a duality between $J_{1x(y)}<1$ and $J_{1x(y)}>1$, the transition point should be exactly at $J_{1x(y)}^{c}=1$.
Otherwise it will generate a contradictory result. Let us explain this point more clear. Suppose $J_{1x}^{c}<1$ for fixed $J_{1y}$ in the straight-line region. Then there are two topological transition points due to the duality, and an intermediate phase exists in the region $J_{1x}^{c}<J_{1x}<1/J_{1x}^c$. Since the VBS with $J_{1x}<J_{1x}^{c}(J_{1x}>1/J_{1x}^{c})$ is adiabatically connected to the $J_{1x}=0 (J_{1x}=\infty)$ topologically nontrivial (trivial) phase, the intermediate phase has contradictory topological properties from the topological transitions of the two different sides, suggesting it should not exist in real system and the transition point is exactly at $J_{1x}^{c}=1$.

In the following sections, we will show that the VBS for $J_{1x}<1$ or $J_{1y}<1$, which is denoted as VBS I, has nontrivial higher-order topological property, while the other one (denoted as VBS II) is topologically trivial.

\begin{figure}[htbp]
\centering \includegraphics[width=9cm]{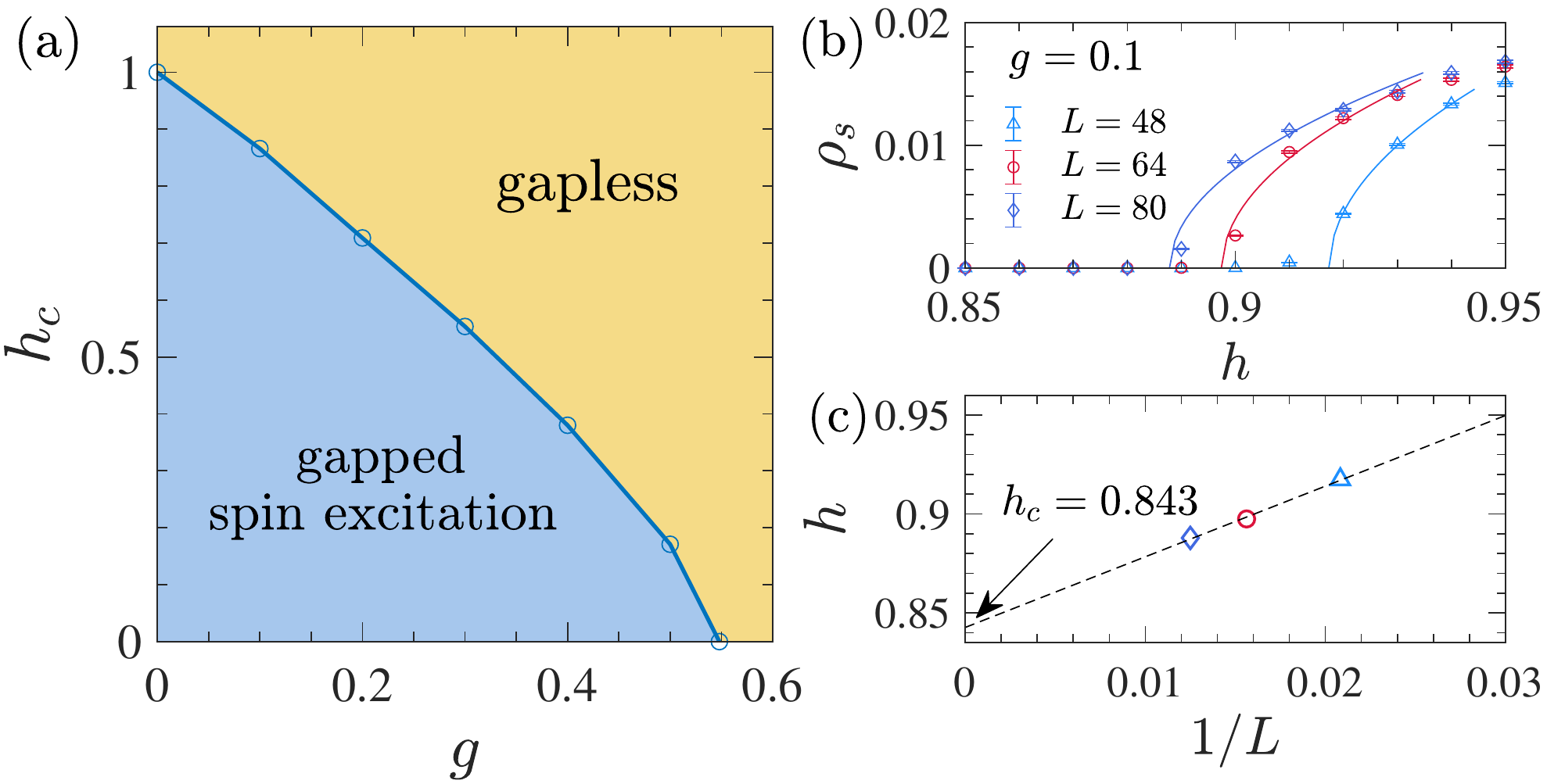} \caption{(a) Phase diagram in the $(g,h_c)$ plane for the Hamiltonian $\hat{H_{0}}+\hat{H_{1}}$ with $\Delta=1$. Below $h_c$, the system remains in the VBS phase with gapped spin excitation. However the spins are aligned along the direction of the applied magnetic field above $h_c$, displaying ferromagnetic behavior. (b) $\rho_s$ vs $h$ near the transition point at $g=0.1$ for several lattice sizes. The data are fitted using the functional form $\rho_s\sim[h-h_c(L)]^{\frac{1}{2}}$ (solid lines). (c) $h_c$ at $g=0.1$ obtained by extrapolating $h_c(L)$ to the thermodynamic limit. Here we consider the $C_4$-symmetric case with $g=J_1/J_2$ and $J_2=1$.}
\label{fig6}
\end{figure}

In the above phase diagrams, the VBS phases are gapped in the spin excitation while the AF state is gapless. To see this explicitly, we add a magnetic field perpendicular to the lattice described by
\begin{eqnarray}\label{eq1}
\hat{H}_1=-\sum_{ {\bf r} }\sum_{i=1}^{4} h S^{z}_{ {\bf r},i }.
\end{eqnarray}
The total Hamiltonian $\hat{H}_0+\hat{H}_1$ is then simulated by the QMC method. The magnetization density
\begin{eqnarray}\label{eq1}
m_z=\frac{1}{N}\sum_{i}^{N}S_i^z,
\end{eqnarray}
is further measured. We find $m_z$ and $\rho_s$ remain to be zero up to a finite magnetic field. Afterwards, they continuously increase, and the system is in a ferromagnetic phase. To determine the transition values precisely, we fit the data with $\rho_s\sim [h-h_c(L)]^{\beta}$ around the critical point for each lattice size\cite{krzakala2008,prokof2008}. As shown in Fig.{\ref{fig6}}(b), our data are compatible with the mean-field exponent $\beta=\frac{1}{2}$. Then $h_c$ is obtained by extrapolating $h_c(L)$ to the thermodynamic limit [see Fig.{\ref{fig6}}(c)]. By this way, the phase diagram in the $(g,h_c)$ plane is plotted in Fig.{\ref{fig6}}(a). Below $h_c$, the system remains in the VBS phase with gapped spin excitations. However the spins are aligned along the direction of the applied magnetic field above $h_c$, displaying a ferromagnetic behavior with gapless spin excitations. The critical value is exactly $h_c=1$ at $g=0$, when the system is decoupled into isolated plaquettes, and can be understood analytically. The spin dimer can be in a spin-singlet ( or spin-triplet) state, with the energy $-\frac{3}{4}J_2$ (or $\frac{1}{4}J_2-h$). So the transition point is $h=J_2=1$, after which the ground state becomes the spin-triplet one where the spins are aligned to the same direction by the magnetic field.

It is interesting to note that the spin-gapped VBS phases are higher-order topological spin states. Next we will demonstrate that gapless spin corner modes will appear on open lattices in both $x$- and $y$-directions due to nontrivial higher-order bulk topology.

\section{the spin corner modes}

\begin{figure}[htbp]
\centering \includegraphics[width=9cm]{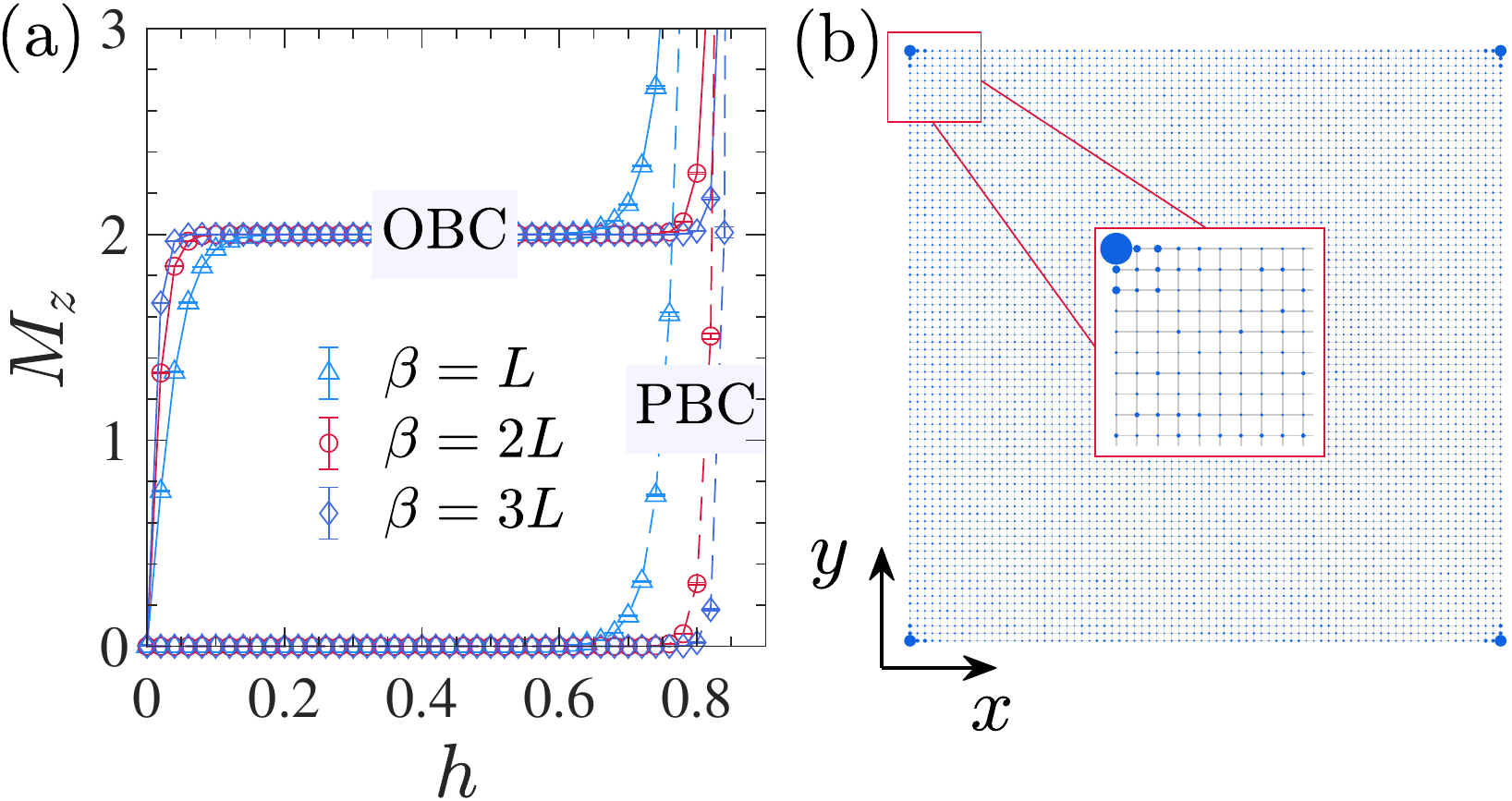} \caption{(a) The total magnetization as a function of $h$ under open and periodic
boundary conditions at different inverse temperatures. The origin of the $M_z=2$ plateau tends to be zero in the limit $\beta\rightarrow \infty (T=0)$.
(b) Local distribution of $M_z$ induced by a magnetic field $h=0.2$ on a $L=40$ open lattice. Four spins are localized around the corners, giving rise to the quantized magnetization $M_z=2$.}
\label{fig7}
\end{figure}

In this section, we demonstrate the spin corner modes, which consists
of the characteristic signatures of nontrivial SHOTPs. Here we consider a square geometry with open boundary condition, which preserves the mirror symmetry $M_x,M_y$, the inversion symmetry, and the rotation symmetry $C_2$. For other geometries, a general principle is that a corner state will appear as long as a domain wall is formed by the two edges crossing at the corner. The SHOTP within $J_{1x(y)}/J_{2}<1$ is adiabatically
connected to the limit $J_{1x(y)}=0$, where there are four dangling
spins at the corners. For finite $J_{1x(y)}$, the gapless spin corner
modes should remain on open lattices in the SHOTP. A infinitesimal
magnetic field $h$ can align these corner spins, thus the total magnetization
becomes $M_{z}=Nm_{z}=2$, compared to $M_{z}=0$ in the bulk SHOTP. Besides, the magnetization is mainly localized near the corners,
and the localization length increases with increasing $J_{1x(y)}$
due to the decreasing of the spin gap. Figure \ref{fig7} shows the
total magnetization as a function of $h$ at $J_{1}/J_{2}=0.1$ (the
$C_{4}$ symmetric case) on a $L=40$ open lattice. It is noted that
the origin of the $M_{z}=2$ plateau tends to be zero as $\beta$
is increased, implying the plateau is quantized as long as $h$ becomes
nonzero in the limit of zero temperature. Beside, the $M_{z}=2$ plateau
persists to a finite $h$, and only begins to vanish when the spin
gap closes. We also show the local distribution of the magnetization
for $J_{1}/J_{2}=0.1$ in Fig.\ref{fig7}(b). Indeed the magnetization
is localized near the corners, implying the gapless corner spins leads
to the quantized magnetization.

Such results are also valid for larger $J_1/J_2$ as long as the ratio is less than the critical value $g_c=0.548$ (see Appendix D). Since the appearance of the spin corner states is a manifestation of the nontrivial higher-order topological property, the
N\'eel-VBS transition is also a higher-order
topological phase transition. Besides, we perform the calculations of $M_z-h$ curves with very small steps near the straight-line boundary of the phase diagram in Fig.5(b). The $M_z$ plateau characterizing the spin corner states immediately vanishes as $J_{1x}$ crosses the boundary located at $J^c_{1x}=1$, thus verifies the straight phase boundary with very high accuracy (see Appendix D).

\section{The higher-order topological invariant}

\begin{figure}[htbp]
\centering \includegraphics[width=6.5cm]{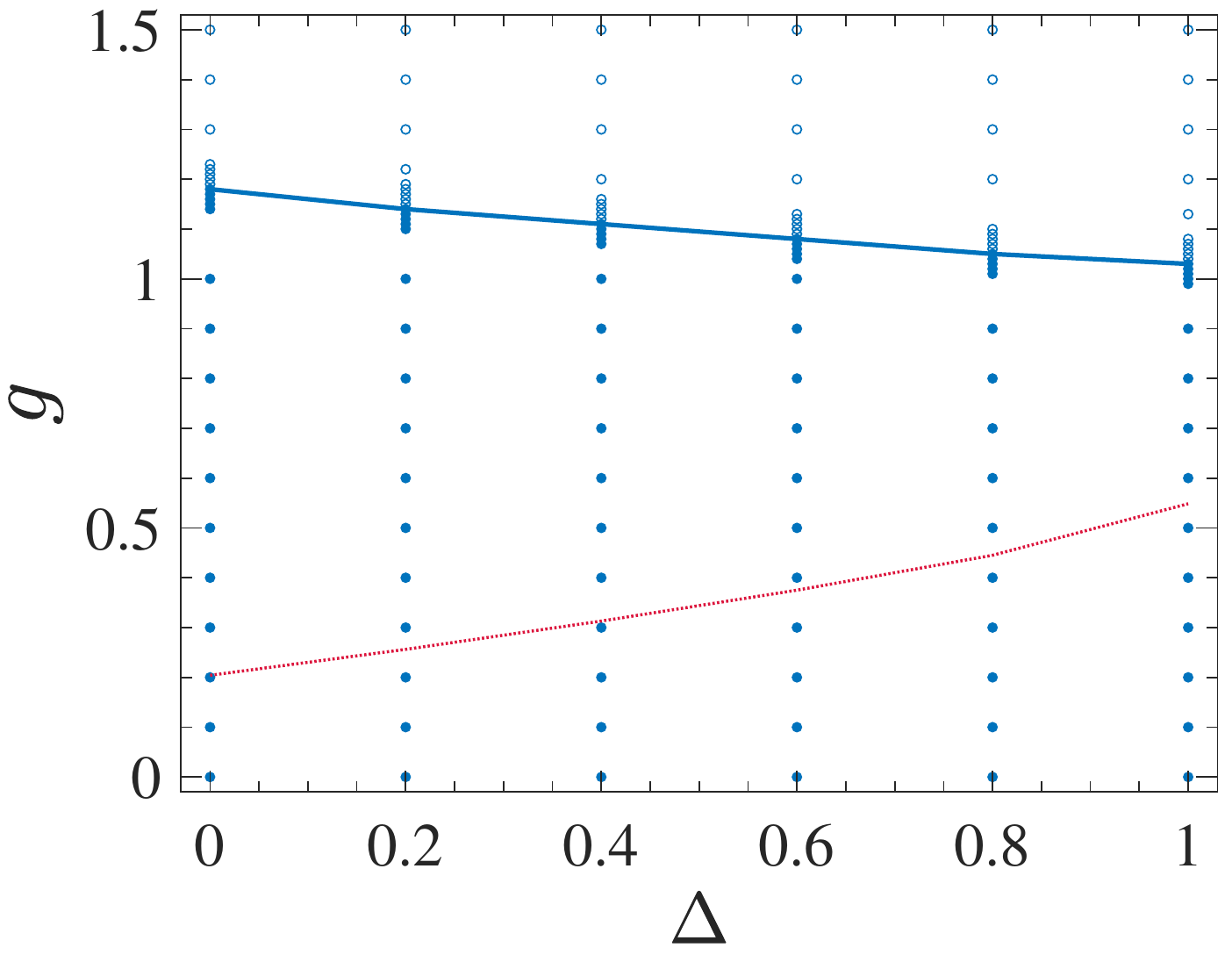} \caption{ The $\mathbb{Z}_4$ Berry phase in the $(g,\Delta)$ plane, where the filled (open) blue circle at each parameter set represent nontrivial value $\pi$ (trivial value $0$). The solid blue line is the phase boundary determined by the $\mathbb{Z}_4$ Berry phase. The QMC result (dotted red line) is also plotted for comparison.}
\label{fig8}
\end{figure}

The appearance of gapless spin corner states is due to the bulk higher-order
topological protection, and the bulk topology is generally characterized
by a topological
invariant. A $\mathbb{Z}_4$ Berry phase is proposed as the topological invariant for the $C_4$-symmetric SHOTP. It is defined with respect to the local twist of the Hamiltonian, which is parameterized by three independent variables $\Theta=(\theta_1, \theta_2, \theta_3)$ ranged in $[0,2\pi)$\cite{araki_2020}.
The Hamiltonian is decomposed into two types of plaquettes with weak and strong bonds, respectively. The twist parameters are introduced to the spin operators on any one of the plaquettes as: $\tilde{S}^{+}_{i}=e^{-i\varphi_i}S^{+}_{i}, \tilde{S}^{-}_{i}=e^{i\varphi_i}S^{-}_{i}$, with $\varphi_i=\sum_{q=1}^{i}\theta_i$ for $i=1,2,3,4$ and $\varphi_4=0$.
The $\mathbb{Z}_4$ Berry phase is defined as
\begin{equation}\label{A3}
\gamma_{z_4}=i\oint_{L_j} \left\langle\Psi_{\Theta}\left|\frac{d}{d \Theta}\right| \Psi_{\Theta}\right\rangle
\end{equation}
where the path $L_j(j=1,2,3,4)$ is in the parameter space: $E_{j-1}\rightarrow G \rightarrow E_j$ with $E_1=(2\pi,0,0), E_2=(0,2\pi,0),E_3=(0,0,2\pi),E_0=E_4=(0,0,0)$, and $G=1/4\sum_{j=1}^{4}E_j$, and $\Psi_{\Theta}$ is the corresponding groundstate many-body wave function at half-filling.
In our calculations, we choose the left-bottom plaquette to apply the local twist (see Fig.1). Figure \ref{fig8} shows the $\mathbb{Z}_4$ Berry phase in the $(g,\Delta)$ plane. As $g$ increases at fixed $\Delta$, the value of the $\mathbb{Z}_4$ Berry phase changes abruptly from $\pi$ to $0$ at a critical $g_c$, marking a topological phase transition. The critical values are $g_c(\Delta=1)=1$ and $g_c(\Delta=1)=1.19$, respectively. For intermediate $\Delta$, the critical value continuously decreases with increasing $\Delta$. Compared to the QMC results, the $\mathbb{Z}_4$ Berry phase greatly overestimates the transition point, and the quantized value persists in the AF phase, which has also been found in other works, and has been attributed to the finite-size error\cite{pollmann2020}. Since the N\'eel-VBS transition is continuous, the two phases coexist near the transition point on small finite lattices. Besides, a finite-size gap exists even in the AF phase. Hence although the $\mathbb{Z}_4$ Berry phase is quantized, it still fails to determine the phase boundary precisely. Another method of calculating the topological
invariant in terms of the many-body quadrupole momentum also suffers significant finite-size errors\cite{wheeler_2019,byungminkang_2019} (see Appendix
E), and identifying a solid many-body topological invariant requires
further studies\cite{yizhiyou2020}.

\section{Connection to the BBH-Hubbard model}

\begin{figure}[htbp]
\centering \includegraphics[width=8.5cm]{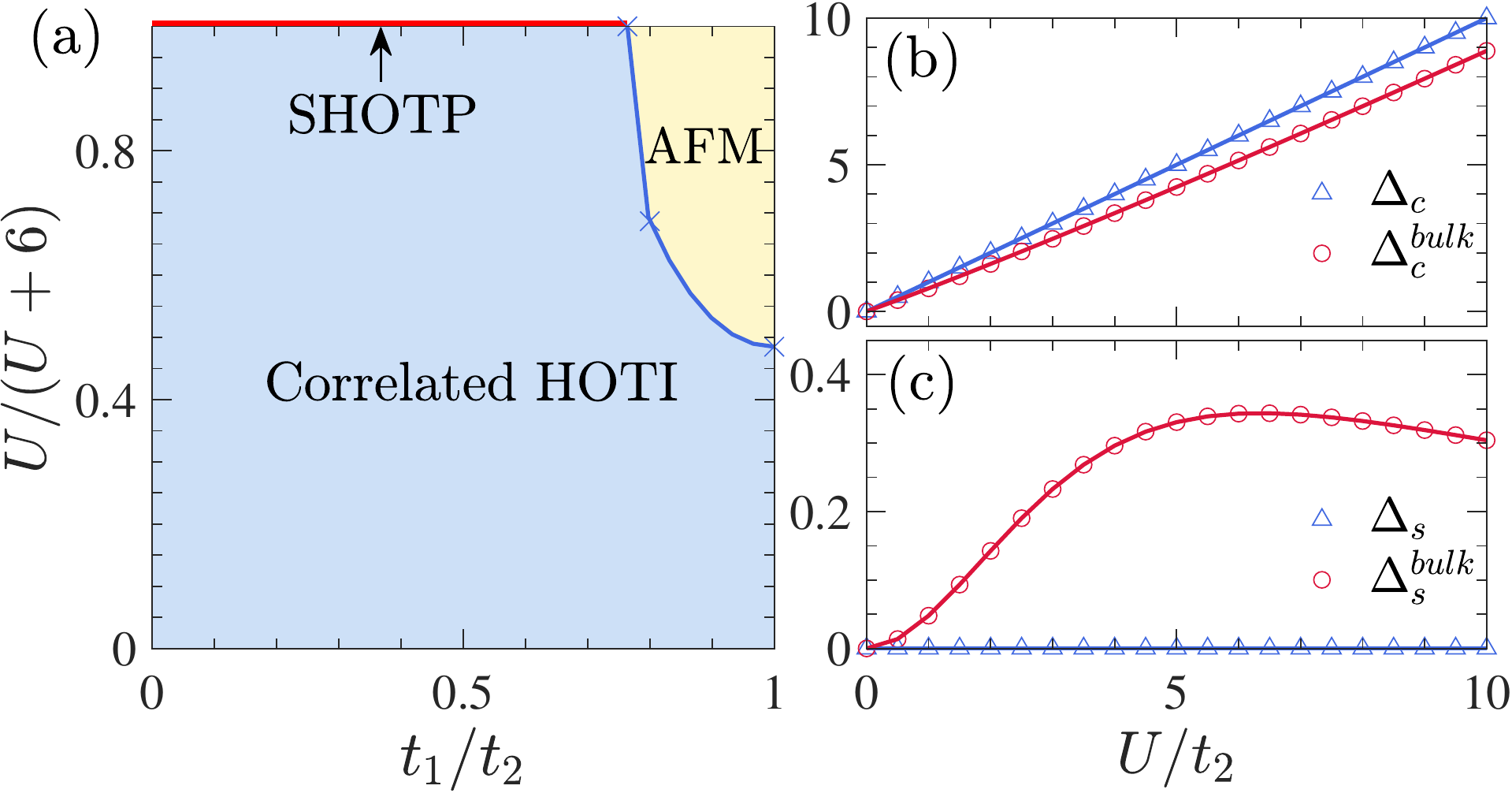} \caption{ (a) Phase diagram of the fermionic BBH-Hubbard model. The $U = \infty$ Heisenberg limit is along the top of the figure, $U/(6 + U) = 1$, and is extracted from the data of Fig. 3. $U_c$ at $t_1/t_2=0.8,1$ are from Refs.\cite{sorella2016,zylu2020}. Charge (b) and spin (c) excitation gaps in the limit $t_1=0$ under periodic and open boundary conditions.}
\label{fig9}
\end{figure}
Here we further compare  SHTOPs discussed in previous sections with correlated HOTIs for a better understanding of their properties.
The Heisenberg model with $\Delta=1$ in Eq.(1) is the large-$U$ limit of the BBH electronic model subjected to an on-site Hubbard interaction $H_U=U\sum_{i}n_{i\uparrow}n_{i\downarrow}$, with $n_{i\sigma}=c^{\dagger}_{i\sigma}c_{i\sigma}$ and $c^{\dagger}_{i\sigma},c_{i\sigma}$ the electronic creation and annihilation operators with spin $\sigma=\uparrow, \downarrow$. At $t_1/t_2=1$, the BBH model becomes the $\pi$-flux lattice hosting 2D Dirac fermions\cite{Wang_2014,Li_2015,assaad2015,sorella2016,hmguo2018}. The Hubbard interaction drives a semimetal-AFM quantum phase transition, with the most accurate value of the critical point $U_c=5.65\pm 0.05$\cite{assaad2015,sorella2016}. In the large-$U$ limit, the relation $J=4t^2/U$ gives the exchange constant in terms of $U$ and the hopping amplitude. Since the VBS-AFM transition happens at $g_c=0.584$, there is a critical ratio for $t_1/t_2$ at $f_c=\sqrt{g_c}=0.764$. Below $f_c$, the correlated HOTI continuously evolves into the SHOTP as $U$ increases, and meanwhile the spin excitation remains gapped in the bulk systems. However above the critical point $f_c$, there is a AFM transition, from which the bulk spin excitation becomes gapless, and hence the correlated HOTI vanishes. The above discussion is summarized as the phase diagram in the $(t_1/t_2, U)$ plane, shown in Fig.\ref{fig9}(a).

In the correlated HOTI, the spin excitation is gapless, but the charge excitation is gapped on open lattices. Figure \ref{fig9}(b) and (c) shows these gaps as a function of $U$ in the limit $t_1=0$. At half filling, each of the four isolated corner sites is occupied by one electron. Adding an opposite spin electron to any such site unavoidably induces the interacting energy $U$, and thus the charge excitation gap is $\Delta_c=U$. However the spin excitation by flipping the corner spin does not cost any energy, and the spin excitation gap is $\Delta_s=0$. We also calculated these gaps in the bulk system, which are defined as follows: $\Delta^{bulk}_c=(E_{N+1,S_z=1/2}+E_{N-1,S_z=-1/2}-2E_{N,S_z=0})/2$, and $\Delta^{bulk}_s=(E_{N,S_z=1}+E_{N,S_z=-1}-2E_{N,S_z=0})/2$, where $E_{N,S_z}$ is the ground-state energy with the total number of electrons $N$ and the total spin in the $z$-direction $S_z$. As shown in Fig.\ref{fig9}(b) and (c), both of them are gapped. So the correlated HOTIs have similar bulk-boundary correspondence as the spin counterparts, i.e., the appearance of gapless spin corner states protected by the higher-order topology.

\section{Conclusions}

In summary, we investigate the higher-order topological properties
of spin analogues of the BBH model using large-scale QMC simulations.
Generally two spin quantum phases, i.e., VBS and AFM, may appear in
the system. A continuous phase transition can be driven by tuning
the ratio between the weak and strong couplings. We identify the VBS
phase with $J_{1x(y)}/J_{2}<1$ to be a nontrivial SHOTP, which is characterized by a bulk spin excitation gap and gapless
spin corner states on open lattice. Besides, in the large-$U$ limit,
the SHOTP is continuously connected to the electronic correlated
HOTI in the BBH-Hubbard model. They share the same bulk-boundary correspondence:
the appearance of gapless corner spin excitations protected by nontrivial
higher-order topology. The SHOTPs also exist in XXZ spin models,
but their regions in the phase diagrams shrink as the anisotropy $\Delta$
decreases. Our results explicitly demonstrate the higher-order topological
properties of spin systems, and contribute to further understandings
of the many-body higher-order topological phenomena.

\section*{Acknowledgments}
The authors thank Xiancong Lu, Zhixiong Li, Tianhe Li, Hiromu Araki, Julian Bibo, Rongqiang He, Hua Jiang, Nvsen Ma for helpful discussions. H.G. acknowledges support from the NSFC grant Nos.~11774019 and 12074022, the Fundamental Research
Funds for the Central Universities and the HPC resources
at Beihang University.
X.Z. and S.F. are supported by the National Key Research and Development Program of China under Grant No. 2016YFA0300304, and NSFC under Grant Nos. 11974051 and 11734002.

\appendix

\section{The analytical form for the spin wave spectrum of the Hamiltonian in Eq.(4)}
The spin wave spectrum of the Hamiltonian in Eq.(4) in the main text can be obtained analytically. The two branches(each of which is two-fold degenerate) are as follows,
\begin{eqnarray*}\label{}
  \omega_1(\mathbf{k}) &=&\sqrt{[C_3 +D({\bf k})-(C_1 \cos k_x +C_2 \cos k_y ) ]/2}, \\ \nonumber
  \omega_2(\mathbf{k}) &=&\sqrt{[C_3 -D({\bf k})-(C_1 \cos k_x +C_2 \cos k_y ) ]/2}, \\ \nonumber
  D({\bf k}) &=&\sqrt{C_4 +D_1({\bf k})+D_2({\bf k})},
\end{eqnarray*}
where $D_1({\bf k})=C_5 \cos \left(k_y \right)+C_6 \cos \left(k_x \right)$, $D_2({\bf k})=4C_1 C_2 \cos \left(k_x \right)\cos \left(k_y \right)$, and
\begin{align}
\nonumber
C_1&=J_{1x} J_2\\ \nonumber
C_2&=J_{1y} J_2\\   \nonumber
C_3&=\left(J_{1x} J_{1y} +J_2^2 +2J_{1x} J_2 +2J_{1y} J_2 \right)\\  \nonumber
C_4&=\left(J_2^4 +J_2^2 J_{1x}^2 +J_2^2 J_{1y}^2 +J_{1x}^2 J_{1y}^2 \right) \\ \nonumber
C_5&=2J_2 J_{1y} \left(J_2^2 +J_{1x}^2 \right)\\ \nonumber
C_6&=2J_2 J_{1x} \left(J_2^2 +J_{1y}^2 \right).
\end{align}

\section{The exact solution of the spin Hamiltonian in Eq.(1) on a plaquette}
In the limit $J_1=0$, the lattice is decoupled into isolated $2\times 2$ plaquettes, on which the spin Hamiltonian in Eq.(1) can be solved by diagonalizing the Hamiltonian matrix. Actually each plaquette is a one-dimensional chain with four sites. For the $S_z=0$ sector, there are six basis states: $| \uparrow\uparrow\downarrow\downarrow\rangle, | \uparrow\downarrow\uparrow\downarrow\rangle, | \uparrow\downarrow\downarrow\uparrow\rangle, | \downarrow\uparrow\uparrow\downarrow\rangle, | \downarrow\uparrow\downarrow\uparrow\rangle, | \downarrow\downarrow\uparrow\uparrow\rangle$. Under the above basis, the Hamiltonian matrix writes as
\begin{align}
H_{S_z=0}=J_2\left(
        \begin{array}{cccccc}
          0 &\frac{1}{2} & 0 & 0 & \frac{1}{2} & 0  \\
          \frac{1}{2} & -\Delta & \frac{1}{2} & \frac{1}{2} & 0 & \frac{1}{2} \\
          0 & \frac{1}{2} & 0 & 0 & \frac{1}{2} & 0 \\
          0 & \frac{1}{2} & 0 & 0 & \frac{1}{2} & 0 \\
          \frac{1}{2} & 0 & \frac{1}{2} & \frac{1}{2} & -\Delta & \frac{1}{2} \\
          0 & \frac{1}{2} & 0 & 0 & \frac{1}{2} & 0 \\
        \end{array}
      \right).
\end{align}
The lowest two eigenvalues are $-J_2(\Delta+\sqrt{\Delta^2+8})/2, -J_2\Delta$, respectively.

For the $S_z=1$ sector, the basis is: $|\uparrow\uparrow\uparrow\downarrow\rangle, |\uparrow\uparrow\downarrow\uparrow\rangle,|\uparrow\downarrow\uparrow\uparrow\rangle, |\downarrow\uparrow\uparrow\uparrow\rangle$, under which the Hamiltonian matrix is,
\begin{align}
H_{S_z=1}=J_2\left(
               \begin{array}{cccc}
                 0 & \frac{1}{2} & 0 & \frac{1}{2} \\
                 \frac{1}{2} & 0 & \frac{1}{2} & 0 \\
                 0 & \frac{1}{2} & 0 & \frac{1}{2} \\
                 \frac{1}{2} & 0 & \frac{1}{2} & 0 \\
               \end{array}
             \right)
.
\end{align}
The lowest eigenvalue is $-J_2$. The $S_z=-1$ sector has exactly the same solution. There is only a diagonal energy for the $S_z=\pm2$ sectors, which is $J_2\Delta$.

Sorting all the above eigenvalues, the ground-state energy is $E_0=-J_2(\Delta+\sqrt{\Delta^2+8})/2$ in the $S_z=0$ sector, and the first-exited state has $E_1=-J_2$ in the $S_z=\pm 1$ sector. Hence a gap with the size $-J_2-E_0$ separates the two lowest levels.

\section{The Zak phase of the 1D fermionic Hamiltonian in Eq.(14)}
\begin{figure}[htbp]
\centering \includegraphics[width=6.5cm]{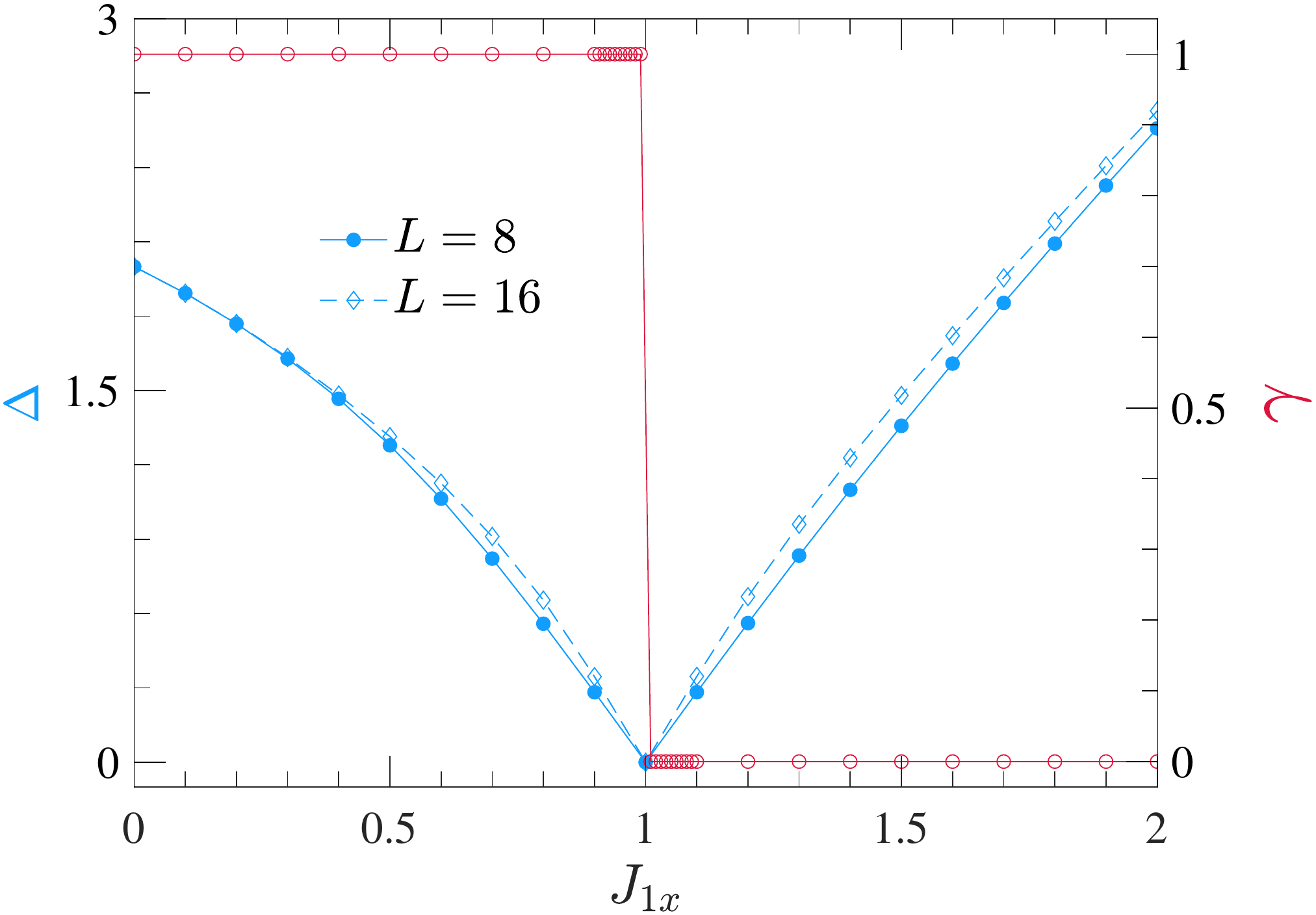} \caption{ The gap between the
ground- and the first-excited states and the Zak phase as a function of $J_{1x}$. The topological phase transition occurs at $J_{1x}=1$, where the gap vanishes, and the Zak phase changes its value from $\pi$ to $0$. Here $J_2=1$.}
\label{afig1}
\end{figure}

We calculate the Zak phase of the ground-state at
half-filling using the twisted boundary conditions\cite{guo2011}. It is defined as
\[
\gamma =\frac{1}{2\pi}\oint i\langle \psi _{\theta }|\frac{d}{d\theta }|\psi _{\theta
}\rangle ,
\]
where $\theta $ is the twisted boundary phase which takes values from $0$ to
$2\pi $ and $\psi _{\theta }$ is the corresponding ground-state many-body
wave function at half-filling obtained by the exact diagonalization method. The $J_{1y}=0$ limit is considered, and the $J_{1x}=0$ limit is the same. As shown in Fig.\ref{afig1}, the gap between the
ground- and the first-excited states vanishes at $J_{1x}=1$, and
the Zak phase $\gamma $ has the value $1$ and $0$ for $J_{1x}<1$ and $J_{1x}>1$, respectively. These clearly show that critical point of the $J_{1y}=0$ limit is exactly at $J_{1x}=1$, determining the on-axis points of the phase diagram in Fig.5(b).

\section{More results of the quantized magnetization by the spin corner states}
\begin{figure}[htbp]
\centering \includegraphics[width=6.5cm]{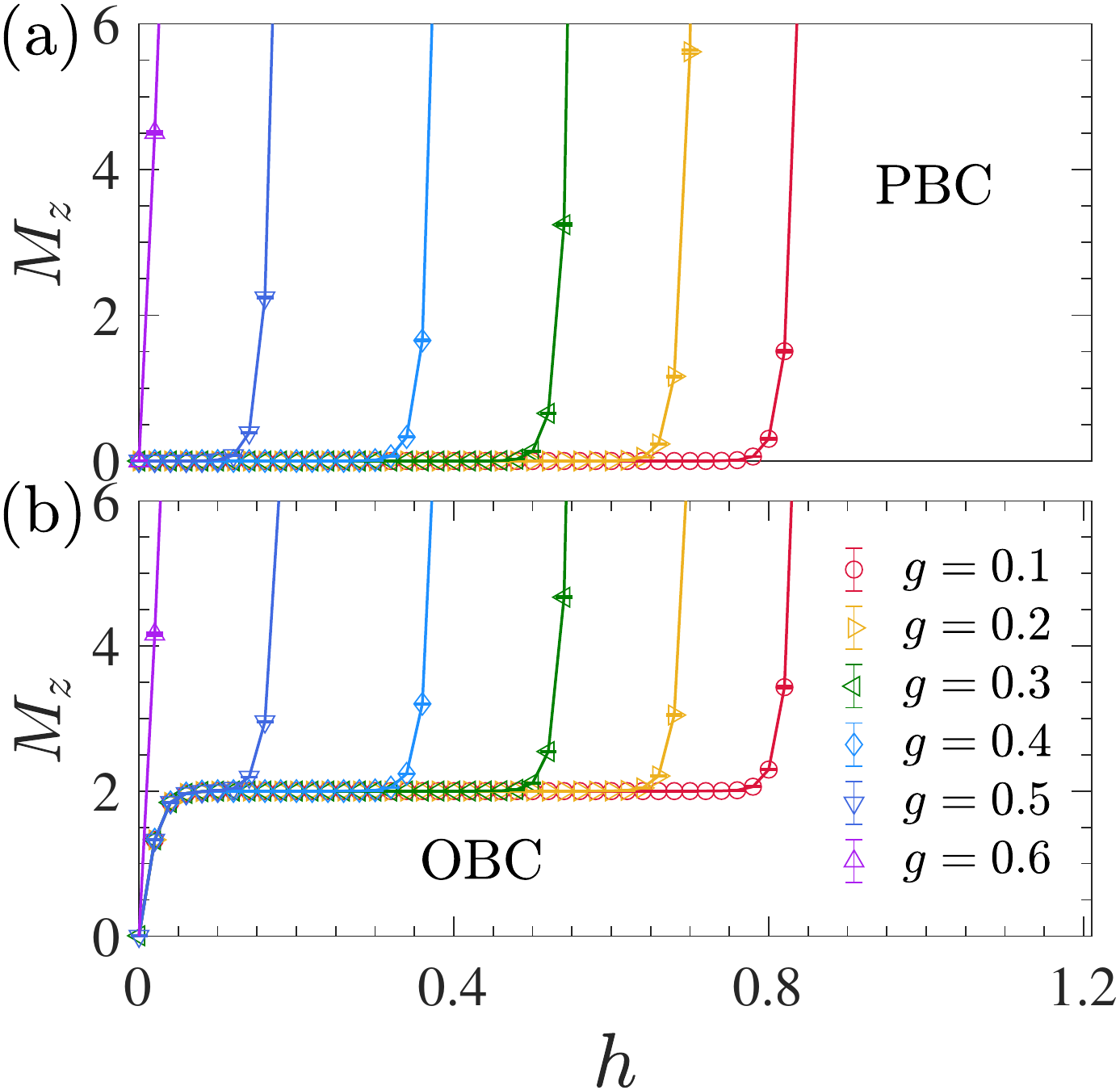} \caption{ The total magnetization as a function of $h$ at different $g$ under periodic (a) and open (b)
boundary conditions. Here the system is $C_4$-symmetric with the lattice size $L=40$ and the inverse temperature is $\beta=2L$.}
\label{afig3}
\end{figure}

Due to the bulk-boundary correspondence, the spin corner modes will appear on the lattices with open boundary condition (OBC) when the system is in the SHOTP. They can be demonstrated by applying an external magnetic field $h$, and a $M_z=2$ plateau, caused by the polarization of the corner spins, exists in the $M_z-h$ curve. Figure \ref{afig3}(a) plots $M_z$ as a function of $h$ on periodic lattices. $M_z$ remains zero up to a finite magnetic field $h_c$, which is proportional to the gap size. When the boundary condition becomes open, the characteristic plateau always exists in the gap for $g<g_c$ (here $g_c=0.548$), but the curves for $g>g_c$ are almost unchanged compared to their periodic counterparts. These clearly show that the appearance of the spin corner states are only below the critical point. Hence a topological phase transition happens accompanying the N\'eel-VBS transition, and the VBS state below $g_c$ has nontrivial spin higher-order topological property.

\begin{figure}[htbp]
\centering \includegraphics[width=6.5cm]{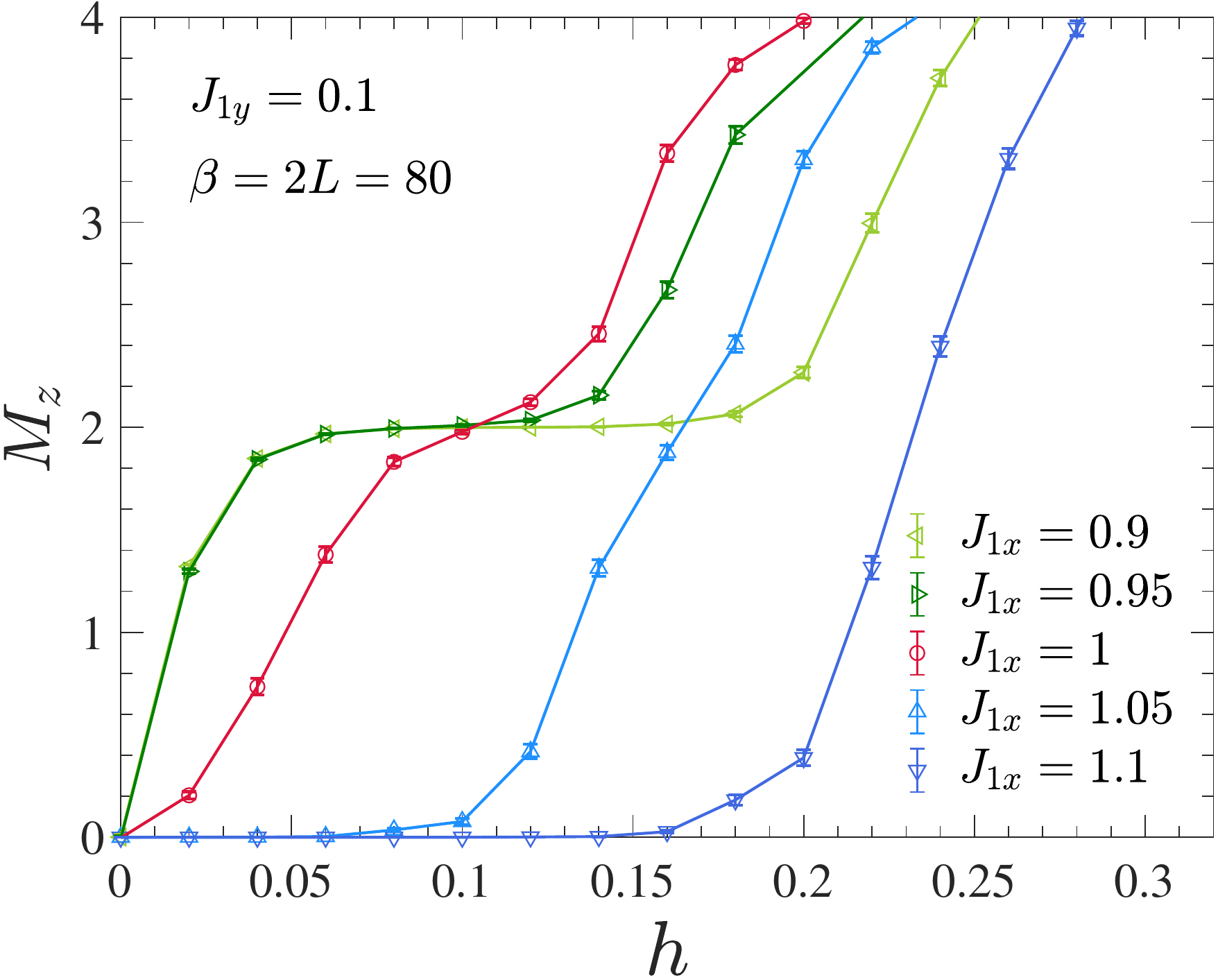} \caption{The total magnetization as a function of $h$ at different $J_{1x}$ with fixed $J_{1y}=0.1$ under open boundary condition. The transition point is at $J_{1x}=1$, thus the quantized plateau immediately vanishes as $J_{1x}>1$. Here the lattice size is $L=40$, and the inverse temperature is $\beta=2L$.}
\label{afig4}
\end{figure}

We also perform similar calculations with very small steps near the straight-line boundary of the phase diagram in Fig.\ref{fig5}(b). As shown in Fig.\ref{afig4}, the $M_z$ plateau immediately vanishes as $J_{1x}$ crosses the boundary located at $J^c_{1x}=1$, verifying the phase boundary with very high accuracy.

\section{Another method of calculating the topological invariant}
Another method of calculating the topological invariant in terms of the many-body quadrupole moment is formulated as follows\cite{wheeler_2019,byungminkang_2019}:
\begin{equation}\label{A1}
q_{xy}=\frac{1}{2\pi}\mathrm{Imlog}\langle\Psi_{G}|\hat{U}_{2}|\Psi_{G}\rangle,
\end{equation}
where $|\Psi_{G}\rangle$ is the many-body ground states, and $\hat{U}_{2}\equiv\exp[i2\pi\hat{q}_{xy}]$
with $\hat{q}_{xy}=\sum_{{\bf r}}xy\hat{n}({\bf r})/(L_{x}L_{y})$
as quadrupole moment density operator per unit cell at position
${\bf r}$. Here $L_{x,y}$ are the sizes of the systems along $x$- and $y$-
directions. The spin-1/2 operators can be mapped to the hardcore boson representation, where the spin operators are written in terms of hardcore boson creation and annihilation operators
\begin{equation}\label{A2}
S^{\dagger}_{i}=b^{\dagger}_i,\ S_{i}=b_i,\ S^{z}_i=b^{\dagger}_i b_i-\frac{1}{2}.
\end{equation}
Hence $\hat{n}({\bf r})$ in $\hat{q}_{xy}$ can be understood as the number operator of hardcore bosons.

\begin{figure}[htbp]
\centering \includegraphics[width=8.5cm]{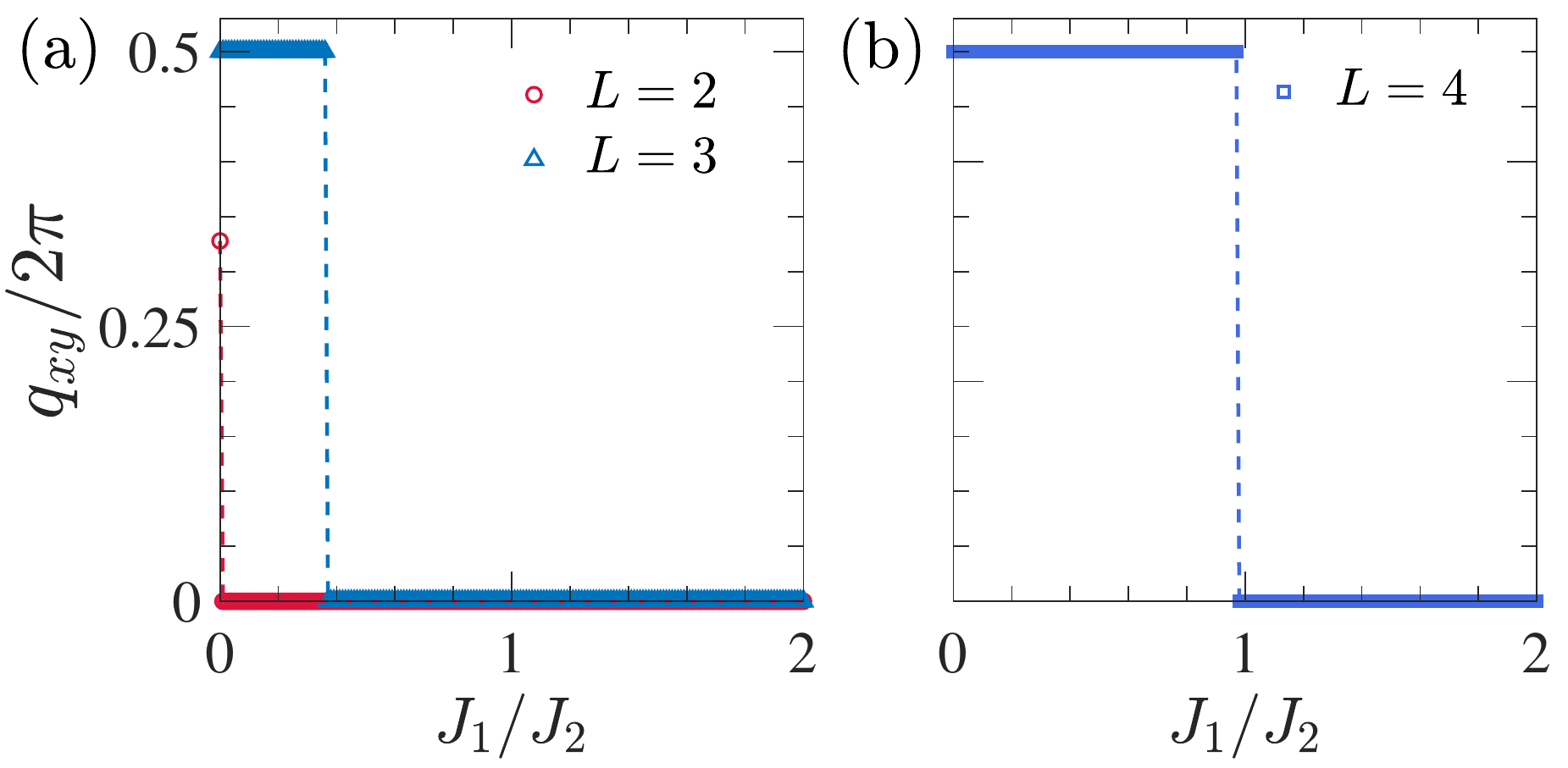} \caption{The size dependence of the quadrupole moment based on the noninteractinng BBH model: (a) $L=2,3$; (c) $L=4$. Here we consider the $C_4$-symmetric case with the critical point at $J_1/J_2=1$. The finite-size error only becomes negligible for $L\geq 4$.}
\label{afig2}
\end{figure}

The definition in Eq.(E1) can be used in the entire parameter space, including the cases with $J_{1x} \neq J_{1y}$. However since the coordinates of each site use
those of the unit cell it belongs to and a 16-site geometry has only $2 \times 2$ unit cells, this
definition has larger finite-size error, and the ED calculations on a 16-site geometry give
uncorrect results. We demonstrate the size dependence of this approach based on the noninteracting BBH model. As shown in Fig.\ref{afig2}, the result becomes reasonable only for $L\geq 4$, i.e., $N\geq 64$ sites, which can hardly be accessed by the ED method.

\bibliography{spin}

\begin{thebibliography}{82}
\bibitem[{nad()}]{nadd}
\bibinfo{note}{In SSE configuration space, the Hamiltonian is written in terms
  of bond operators, which are further decomposed into diagnol and off-diagnol
  ones. Then the powers of the Hamiltonian can be expressed as sums of products
  of these bond operators. In the simulation, we have a sequence of these
  operators: $\prod_{i=1}^{n} H_{a_{i}, b_{i}}$, where $a_{i} \in\{1,2\}$
  corresponds to the type of operator $(1,$ diagonal; 2, off-diagonal) and
  $b_{i} \in\left\{1, \ldots, N_{b}\right\}$ is the bond index. Specificly, the
  off-diagonal operator acting on the nearest-neighbor pair $(i,j)$ is $H_{2,
  b}=\frac{1}{2}\left[S_{i(b)}^{+} S_{j(b)}^{-}+S_{i(b)}^{-}
  S_{j(b)}^{+}\right]$. $N_{\alpha}^{+}\left(N_{\alpha}^{-}\right)$ count the
  total number of operators transporting spin in the positive (negative)
  direction. For example: supposing $(i,j)$ is along the positive $\alpha$
  direction and $S_{i(b)}^{+} S_{j(b)}^{-}$ in $H_{2, b}$ takes effect, the
  spin transports along the positive $\alpha$ direction, and $N^{+}_{\alpha}$
  counts once.}
\end{thebibliography}

\end{document}